\documentclass[12pt]{elsart}
\usepackage{epsf}

\newcommand\eq[1] {(\ref{#1})}

\newcommand{\nonum}{\nonumber \\}
\newcommand{\beqa}{\begin{eqnarray}}
\newcommand{\eeqa}[1]{\label{#1}\end{eqnarray}}
\newcommand{\beq}{\begin{equation}}
\newcommand{\eeq}[1]{\label{#1}\end{equation}}

\newcommand{\Imag}{\mathop{\rm Im}\nolimits}

\newcommand{\Md}{\partial}

\newcommand{\Ga}{\alpha}

\newcommand{\Gd}{\delta}

\newcommand{\Gve}{\varepsilon}

\newcommand{\Gl}{\lambda}

\newcommand{\Gm}{\mu}

\newcommand{\Go}{\omega}

\def\BE{{\bf E}}

\def\BH{{\bf H}}

\begin{document}
\begin{frontmatter}

\title{Opaque perfect lenses}
\author{Graeme W. Milton\thanksref{Milton}}
\address{Department of Mathematics, University of Utah,
Salt Lake City UT 84112 USA}
\author{Nicolae-Alexandru P. Nicorovici and  Ross C. McPhedran}
\address{ARC Centre of Excellence for Ultrahigh-bandwidth Devices for
Optical Systems (CUDOS)\\
School of Physics, University of Sydney, Sydney NSW 2006 Australia}
\thanks[Milton]{\tt email milton@math.utah.edu}
\begin{abstract}
The response of the ``perfect lens'', consisting of a slab of lossless material
of thickness $d$ with $\Gve_s=\Gm_s=-1$ at one frequency $\Go_0$ is investigated.
It is shown that as time progresses the lens becomes increasingly opaque
to any physical TM line dipole source located at a distance $d_0<d/2$ from the lens and
which has been turned on at time $t=0$. Here a physical source is defined as one which supplies a
bounded amount of energy per unit time. In fact the lens cloaks the source so that it is
not visible from behind the lens either. For sources which are turned on exponentially
slowly there is an exact correspondence between the response
of the perfect lens in the long time constant limit
and the response of lossy lenses in
the low loss limit. Contrary to the usual picture where the field intensity
has a minimum at the front interface we find that the field diverges to
infinity there in the long time constant limit.

\end{abstract}

\end{frontmatter}

\section{Introduction}
\setcounter{equation}{0}
Recently there has been growing interest in superresolution,
i.e. the fact that an image can be sharper than the wavelength of the radiation,
which is in direct contrast to the proof of Abbe in 1873 that the resolution
of a normal lens is at most about $\Gl/(2n)$ where $\Gl$ is the
wavelength and $n$ is the refractive index. Initial progress towards superresolution
had the characteristic feature that the decrease in spot size
was accompanied by an unwanted sharp decrease in the intensity at the center of the spot relative to that of the Airy pattern
(with an accompanying increase in sidelobe intensity): see \cite{Sales:1997:FLO} and references therein. Although its
significance was not recognized at the time, a breakthrough came in 1994 when
it was discovered that quasistatic line sources could have arbitrarily sharp images.
Specifically it was found \cite{Nicorovici:1994:ODP} that a coated cylinder
with inner and outer radii $r_c$ and $r_s$ and having a real core dielectric
constant $\Gve_c$, a shell dielectric constant $\Gve_s$ close to $-1$ (with a
small positive imaginary part) and a matrix dielectric constant $\Gve_m=1$
would have some rather strange properties in the quasistatic limit (where the
free-space wavelength is infinitely long compared to the structure).
In particular a line source aligned with the cylinder axis and positioned
outside the cylinder radius at a radius $r_0$ with $r_*<r_0<r_*^2/r_s$
where $r_*=r_s^2/r_c$ would have an arbitrarily sharp image positioned
at a radius $r_*^2/r_0$ outside the coated cylinder. This image would
only be apparent beyond the radius  $r_*^2/r_0$;
closer to the coated cylinder the potential was numerically
found to exhibit enormous
oscillations. The reason that one finds an image at this radius is
that it was shown that the effect of the shell was to magnify the
core, so it was equivalent to a solid cylinder of radius $r_*$.
By the method of images in two-dimensional electrostatics
the field outside the equivalent solid cylinder is that due
to the actual source plus an image source at the radius $r_*^2/r_0$.
However in contrast to electrostatics, the image source now lies
in the physical region outside the coated cylinder. The paper
\cite{Milton:2005:PSQ}
contains an in depth review of the results of the 1994 paper,
correcting some minor errors.

In independent work, Pendry \cite{Pendry:2000:NRM} claimed that
line sources could have arbitrarily sharp images, even beyond
the quasistatic regime, and he realized the deep significance of this
result for imaging. His analysis suggested that the Veselago lens,
consisting of a slab of material having thickness $d$, relative electric
permittivity $\Gve_s=-1$, and relative magnetic permeability $\Gm_s=-1$
(and thus having a refractive index $n=-1$) would act as a superlens perfectly
imaging the fields near the lens and shifting them by the distance
$2d$. Basically each interface of the lens acts like a mirror: Maxwell's
equations are satisfied when the $\BE$ and $\BH$ fields on opposite sides
of each interface are reflections of each other, and the two reflections
about the two interfaces give a net translation of the fields by the distance $2d$.
There were some flaws in Pendry's original analysis. In particular a point source
at a distance $d_0<d$ from the lens, could not have an actual point source as
its image, since this would imply a singularity in the fields at the image point
which cannot happen \cite{Maystre:2004:PLM}.
In fact there is no time harmonic solution in this
case \cite{Garcia:2002:LHM,Pokrovsky:2002:DLH}
since surface polaritons of vanishingly small wavelengths
cause divergences \cite{Haldane:2002:ESM}. While experiment
has provided evidence for superresolution
\cite{Lagarkov:2004:NPI,Grbic:2004:ODL,Fang:2005:SDL,Melville:2005:SRI,Korobkin:2006:ENF},
to make theoretical sense of Pendry's claim one has to regularize the
problem, say by making the slab lens slightly lossy or by
switching on the source for a finite time. A careful analysis of
the lossy case was made in \cite{Shvets:2003:ASP,Podolskiy:2005:NSS}, and a rigorous mathematical
proof of superlensing for quasistatic fields was given
in \cite{Milton:2005:PSQ} (see also \cite{Merlin:2004:ASA} where a careful time harmonic analysis was given for real
$\Gve_s$ and $\Gm_s$ close, but not equal to $-1$). Both for the quasistatic case
\cite{Milton:2005:PSQ} and for the full time harmonic Maxwell equations \cite{Milton:2006:CEA,Yaghjian:2006:PWS}
it was shown that contrary to the conventional
explanation where the field intensity has a minimum at the
front interface of the lens, the field actually diverges to
infinity in two anomalously resonant layers of width $2(d-d_0)$, one centered on the
front interface and one centered on the back interface. Indications
of large fields in front of the lens \cite{Merlin:2004:ASA,Rao:2003:AEW,Shvets:2003:PAM,Cummer:2003:SCS,Guenneau:2005:PCR}
were followed by definitive numerical evidence of enormous fields \cite{Podolskiy:2005:OSM}.
When $d_0<d/2$
the resonant layers interfere with the source. It was discovered
\cite{Milton:2006:CEA}
(following a suggestion of Alexei Efros that the energy
absorbed by the lens may be infinite), that finite energy point
or line sources or polarizable point or line dipoles less than
a distance $d/2$ from the lens become cloaked, and are
essentially invisible from outside the distance $d/2$
from the lens. Thus the Vesalago lens, in the limit
as the loss tends to zero does not perfectly
image physical sources that lie closer to the lens than
a distance $d/2$.

The following is a quick back of the envelope explanation of cloaking
of a single polarizable dipole
due to anomalous localized resonance (see the paper \cite{Milton:2006:CEA}  for more details).
First we should point that anomalous localized resonance is
a phenomenon where as the loss in the system goes to zero the
fields diverge to infinity within a specific region (the region of anomalous resonance) not associated
with a change in character in that region of the underlying partial
differential equation, but the fields approach smooth fields outside that region \cite{Milton:2005:PSQ}.
Now consider a polarizable (line or point) dipole outside the Veselago superlens
surrounded by fixed sources. Let $E$ denote the electric field at the dipole
in the absence of the dipole. Let $E_r$ denote the field acting on the
dipole, when its dipole moment is $k$ and the surrounding fixed sources
are absent (but the superlens is still present). Due to linearity
one has that $E_r=c(\Gd)k$, where $\Gd$ is a parameter measuring
the moduli governing the loss in the superlens with $\Gd=0$
corresponding to a loss-less lens, and $c(\Gd)$ is in general
tensorial but for simplicity of argument let us suppose that it
is scalar valued. Due to anomalous resonance $|c(\Gd)|\to\infty$
as $\Gd\to 0$ when the dipole is within a distance $d/2$ from
the superlens (and approaches zero otherwise).
Now by superposition the total field acting on
the dipole will be $E_t=E+E_r$, and the induced dipole moment
will be $k=\Ga E_t$ where $\Ga$ is the polarizability of the
dipole. So we have that $k=\Ga[E+c(\Gd)k]$ and solving for
$k$ gives $k=\Ga_*E$ where the ``effective polarizability''
\beq \Ga_*=\frac{\Ga}{1-\Ga c(\Gd)}
\eeq{C.1}
goes to zero (and is asymptotically almost independent of $\Ga$)
when the dipole is positioned within the cloaking region $0<d_0<d/2$
where  $|c(\Gd)|\to\infty$. Basically the polarizable
dipole causes the superlens to build up a localized resonance
and the fields from this resonance almost exactly cancel the
field that would otherwise act on the polarizable line dipole: in
effect, the polarizable line dipole feels a vanishingly small
field. So its induced dipole moment is close to zero and consequently
it perturbs the field only slightly outside the resonant region.
It is effectively invisible. This
simple argument has to be replaced by more sophisticated
(energy based) arguments to establish the cloaking of
collections of dipoles. In contrast
to superlensing, which requires very low loss materials to
get a significant enhancement of resolution \cite{Podolskiy:2005:NSS,Yaghjian:2006:PWS}
cloaking effects occur for materials which have moderate loss: see figure 3
in \cite{Milton:2006:CEA}.

      The hope has persisted that a source turned on at
time $t=0$ would be perfectly imaged by a lossless Veselago lens
(the perfect lens) as $t\to\infty$. This was first suggested by
G\'{o}mez-Santos\cite{Gomez:2003:UFT} and subsequently Yaghjian and
Hansen\cite{Yaghjian:2006:PWS} gave a detailed analysis. Both papers
took into account the fact that due to dispersion $\Gm_s(\Go)$ and
$\Gve_s(\Go)$ can only equal $-1$ at one frequency $\Go_0$. At
nearby frequencies one has \beqa
\Gve_s(\Go)=-1+a_\Gve(\Go-\Go_0)+O[(\Go-\Go_0)^2],\nonum
\Gm_s(\Go)=-1+a_\Gm(\Go-\Go_0)+O[(\Go-\Go_0)^2], \eeqa{1.0} where,
due to causality, the dispersion coefficients (with
$\Gve_s(\Go_0)=\Gm_s(\Go_0)=-1$) necessarily satisfy the
inequalities \cite{Yaghjian:2005:IBA,Yaghjian:2006:PWS}
\beq a_\Gve=\left.\frac{d
\Gve_s}{d \Go}\right|_{\Go=\Go_0}\geq\frac{4}{\Go_0},\quad \quad
a_\Gm=\left.\frac{d \Gm_s}{d
\Go}\right|_{\Go=\Go_0}\geq\frac{4}{\Go_0}, \eeq{1.1a} which force
them to be positive.  [These inequalities are also a corollary
of bounds derived in \cite{Milton:1997:FFR}. To see this, suppose $\Go$ and
$\Go_0$ belong to a frequency interval where $\Gve_s$ is real,
and that we seek bounds which correlate the values that $\Gve_s(\Go)$
and $\Gve_s(\Go_0)$ can take. Without loss of generality let us suppose
that $\Go^2>\Go_0^2$. Then from equation (6) in \cite{Milton:1997:FFR}
we have the sharp bound
\beq \Gve_s(\Go)-1\geq \max\{\Gve_s(\Go_0)-1,\Go_0^2[\Gve_s(\Go_0)-1]/\Go^2\}
\eeq{1.1b}
which when $\Gve_s(\Go_0)=-1$ reduces to
\beq \Gve_s(\Go)\geq 1-2\Go_0^2/\Go^2. \eeq{1.1c}
By substituting the Taylor expansion \eq{1.0} in this inequality
and letting $\Go\to\Go_0$ we obtain the first inequality in \eq{1.1a}.
The second inequality is obtained by similar arguments applied to
the magnetic permeability.]

For simplicity it is assumed that the surrounding matrix material has
$\Gm_m=\Gve_m=1$ for all frequencies. It was
shown in the papers \cite{Yaghjian:2006:PWS,Gomez:2003:UFT}
that the field at any given time would be finite
except at the source. Also figure 1 in \cite{Gomez:2003:UFT} shows the field has a
local intensity minimum at the front interface and
it was claimed in \cite{Yaghjian:2006:PWS} that as $t\to\infty$
the field would diverge only in a single layer of width $2(d-d_0)$,
centered on the back interface. However, here we will show that,
again contrary to the conventional picture,
the situation is precisely analogous to what occurs in a lossy
lens as the loss goes to zero. The field also diverges to infinity
in the layer of width  $2(d-d_0)$ centered on the front interface,
and as a consequence cloaking occurs when the source is less than
a distance $d/2$ from the lens. The image of a constant energy
source in this cloaking region becomes rapidly dimmer and dimmer
as time increases. So instead of the lens being perfect, it is
actually opaque to such sources, and cloaks them: not only
is the source dim behind the lens, it is also dim in front
of the lens. Essentially all of the energy produced by the
source gets funneled into the resonant regions which
continually build up in intensity. Thus the
claim \cite{Gomez:2003:UFT}
that ``{\it even within the self-imposed idealizations
of a lossless (for $\Go=\Go_0$) and purely homogeneous,
left handed material, Pendry's perfect lens proposal is
correct}'' has to be qualified. It is {\it only true} for physical sources
located further than a distance $d/2$ from the lens. For physical
sources located less than a distance $d/2$ from the lens
the image is completely different from what
would appear if the lens were absent because the source
interacts with the resonant fields in front of the lens.
Although our analysis assumes a point source, any source
of finite extent can be viewed as a superposition of
dipolar sources and will create fields in front of the
lens that interact with the source.

\section{Analysis}
\setcounter{equation}{0}

        Simple energy considerations indicate that something
strange must happen when $d_0<d/2$.
From equation (62) in \cite{Yaghjian:2006:PWS} we see that a
source of constant strength $E_0$ switched on at $t=0$ creates an electric
field which near the back interface (and outside the lens) scales approximately as
\beq E \sim E_0 t^{1-d_0/d}.
\eeq{1.1}
The stored electrical energy $S_E(t)$ will scale as the square of this, and consequently
the time derivative of the stored electrical energy will scale approximately as
\beq \frac{dS_E}{dt}\sim E_0^2t^{1-2d_0/d},
\eeq{1.2}
which blows up to infinity as $t\to \infty$. If the source produces a
bounded amount of energy per unit time we have a contradiction. The
conclusion is that if the energy production rate of the source is
bounded then necessarily $E_0$ must decrease to zero as $t\to \infty$.
(If it approached any other equilibrium value then again
we would have a contradiction). This sounds rather paradoxical
but it could be explained if there was a resonant region in
front of the lens, creating a sort of optical molasses, requiring
ever increasing amounts of work to maintain the constant strength
$E_0$.

       Let us see that there is a resonant region in
front of the lens through an adiabatic treatment of the problem.
For simplicity we assume a TM line dipole source located along the $Z$-axis (which we
capitalize to avoid confusion with $z=x+iy$) and that the
slab faces are located at the planes $x=d_0$ and $x=d_0+d$.
Instead of assuming that the source is turned on sharply
at $t=0$ and thereafter remains constant we assume that it has been turned on exponentially slowly
beginning in the infinite past.
The source generates a field with the plane wave expansion
\beq H_Z^{\rm dip}(x,y,t)=
\int_{-\infty}^{\infty}dk_y~ a(k_y)e^{i(k_x x+k_y y-\Go t)}
~~{\rm with}~~ k_x=\sqrt{\Go^2/c^2-k_y^2},
\eeq{1.3}
for $x>0$, which interacts with the lens, where the coefficients $a(k_y)$ need to be determined
and the square root in \eq{1.3} is chosen so $\Imag k_x>0$ to ensure that the
waves due to the source decay as $x$ increases. The frequency
\beq \Go=\Go_0+i/T \eeq{1.3a}
is complex and $T$  is a measure of the time the source has been ``switched on'' until time
$t=0$. It does not make sense to analyse this model in the limit as $t\to \infty$
since everything diverges exponentially in that limit. Rather we consider the model at time $t=0$
at which point the source has been approximately constant for
a very long period of time of the order of $T$. Thus investigating the asymptotic
behavior as $T\to\infty$ at $t=0$ in this model is analogous
to investigating the asymptotic behavior as $t\to\infty$ of a constant amplitude
source which has been switched on at time $t=0$.

For a
dipole line source we have
\beq H_Z^{\rm dip}(x,y,t)=\frac{\pi\Go_0 e^{-i\Go t}}{2}\left(-k^{\rm o}\frac{\Md}{\Md x}+ik^{\rm e}\frac{\Md}{\Md y}\right)
H^{(1)}_0\left((\Go/c)\sqrt{x^2+y^2}\right),
\eeq{1.3b}
in which $H^{(1)}_0$ is a Hankel function of the first kind
and $k^{\rm e}$ is the (possibly complex) strength at $t=0$ of the dipole component which has
an associated electric field with even symmetry about the $x$ axis
and $k^{\rm o}$ is the (possibly complex) strength at $t=0$
of the dipole component which has an associated electric field with odd
symmetry about the $x$ axis: these dipole strengths have been
normalized to agree with the definitions in \cite{Milton:2005:PSQ} and \cite{Milton:2006:CEA}.
By substituting the plane
wave expansion [see formula (2.2.11) in \cite{Chew:1995:WFI}]
\beq H^{(1)}_0\left((\Go/c)\sqrt{x^2+y^2}\right)=\frac{1}{\pi}\int_{-\infty}^{\infty}dk_y~\frac{e^{i(k_x x+k_y y)}}{k_x},
\eeq{1.3c}
with
\beq k_x=\sqrt{\Go^2/c^2-k_y^2}, \eeq{1.3d}
in \eq{1.3c} we see that
\beq  a(k_y)=-\Go_0[k^{\rm e}(k_y/k_x)+ik^{\rm o}]/2.
\eeq{1.4}

       We look for a particular solution of Maxwell's equations
where all the fields, and not only the source, vary with time
as $e^{-i\Go t}$ where $\Go$ is given by
\eq{1.3a}. This solution is obtained by substituting this complex
value of $\Go$ into the time harmonic Maxwell's equations. Specifically
with $\Go=\Go_0+i/T$ and with the lens having the least
possible dispersion, $\Gve_s$ and $\Gm_s$ will according to
\eq{1.0} have the complex values
\beq \Gve_s=-1+ia_{\Gve}/T+ O(1/T^2), \quad
 \Gm_s=-1+ia_{\Gm}/T+ O(1/T^2),
\eeq{1.5}
In other words, apart from the modulating factor of $e^{-i\Go t}$, the
mathematical solution for the fields is exactly the same
as for a lossy material with $\Gm''_s$ and $\Gve''_s$ approximately proportional to $1/T$
for large $T$. A correspondence of this sort was noted before \cite{Yaghjian:2006:PWS}
but not fully exploited. By this argument it immediately follows that for fixed $k^{\rm e}$ and $k^{\rm o}$ the fields will diverge as
$T\to\infty$ in {\it two} possibly overlapping layers of the same width $2(d-d_0)$ one
centered on the back interface and one centered on the front interface. In particular, in front of the lens,
with $2d_0-d<x<d_0$, equations (4.18) and (4.19) of \cite{Milton:2006:CEA} imply
\beqa H_Z(x,y,t)&\approx & H_Z^{\rm dip}(x,y,t)\nonum &~& -\Go_0e^{-i\Go t}
\{[g^{\rm e}(z)-g^{\rm e}(\bar{z})]/2+[g^{\rm o}(z)+g^{\rm o}(\bar{z})]/(2i)\},
\eeqa{1.6}
where $z=x+iy$, $\bar{z}=x-iy$ and
\beq g^{\rm p}(z)=-iqk^{\rm p}[a_{\Gve}/(2T)]^{(2d_0-d-z)/d}Q_0(2d-2d_0+z),
\eeq{1.7}
with
\beq
Q_0(b)=\frac{\pi}{2d\sin[\pi b/(2d)]},
\eeq{1.7a}
in which $q=1$ for p$=$e and $q=-1$ for p$=$o. Thus we see that $g^{\rm p}(z)$
and hence $H_Z(x,y,t)$ diverges as $T\to\infty$ within a distance $d-d_0$ from
the front of the lens. When $d_0<d/2$ this resonant region interacts with
the source creating the ``optical molasses'' that we mentioned. We have
not done the computation, but presumably if one took $k^{\rm o}=0$ and chose
$k^{\rm e}$ to depend on $T$ in such a way that the source produces a given
($T$ independent) amount of energy at time $t=0$ then one would find
as $T\to\infty$ that the field would be localized and resonant
in two layers of width $d$ which touch at the slab center. We remark that
such field localization was found in the quasistatic case in the low loss
limit \cite{Milton:2006:CEA} and also when two opposing sources are
placed a distance $d/2$ behind and in front of the lens
\cite{Cui:2005:LEE,Boardman:2006:NRC}

      We only considered a particular solution to the equations. The general
solution is of course the sum of a particular solution plus a solution
to the homogeneous equations with no sources present, which we call a
resonant solution. Since the lens is lossless, energy must be conserved
and so a resonant solution which is zero and has zero total energy
in the infinite past, must be zero for all time. Therefore the particular
solution we considered is the only solution which satisfies the boundary
condition of being zero in the infinite past.

       No immediately apparent problems occur for
line sources with $d_0$ between $d/2$ and $d$.
While the stored electrical energy $S_E(t)$ in the resonant regions
increases without bound, we see from \eq{1.2}
that the rate of increase diminishes with time. Similarly the rate of
increase of magnetic energy diminishes with time. Therefore the image of
such sources will get brighter and brighter as $t\to\infty$ approaching the
same brightness as the original source without the lens present.
However because the energy stored in the resonant regions is so
large it may be the case that slight variations in the
intensity of the source or slight non-linearities or
slight inhomogeneities in the permeability and permittivity of the
lens will scatter radiation and destroy the ``perfect image''. The
spatial dispersion of the dielectric
response of the slab will also limit resolution \cite{Larkin:2005:IPL}.
Finally we remark that we have assumed that the radiation coming
from the source is coherent.

\section*{Acknowledgements}
        The authors thank Alexei Efros for helpful comments on the
manuscript and for suggesting that cloaking
may be a feature of perfect lenses, and not just of lossy lenses
in the low loss limit. The referees are thanked for their remarks
which improved the paper.
G.W.M. is grateful for support from the
National Science Foundation through grant DMS-0411035,
and from the Australian Research Council.
The work of N.A.N. and R.C.McP.
was produced with the assistance of the Australian Research
Council.

\bibliography{/u/ma/milton/tcbook,/u/ma/milton/newref}

\end{document}